\documentclass[9pt,conference]{IEEEtran}

\usepackage{waspaa25} 

\usepackage{bm} 
\usepackage{color, colortbl, arydshln}
\usepackage{graphicx}
\usepackage{subcaption}  
\usepackage[margin=1in]{geometry} 

\title{Real-time speech enhancement in noise for throat microphone using neural audio codec as foundation model}

\name{Julien Hauret$^{1}$,
      Thomas Joubaud$^{2}$,
      Éric Bavu$^{1}$}

\address{$^{1}$LMSSC, Conservatoire national des arts et métiers, Paris, France \\
$^{2}$Acoustics and Protection of the Soldier, French-German Research Institute of Saint-Louis, France}

\begin{document}

\maketitle

\begin{abstract}
We present a real-time speech enhancement demo using speech captured with a throat microphone. This demo aims to showcase the complete pipeline, from recording to deep learning-based post-processing, for speech captured in noisy environments with a body-conducted microphone. The throat microphone records skin vibrations, which naturally attenuate external noise, but this robustness comes at the cost of reduced audio bandwidth. To address this challenge, we fine-tune Kyutai's Mimi—a neural audio codec supporting real-time inference—on Vibravox, a dataset containing paired air-conducted and throat microphone recordings. We compare this enhancement strategy against state-of-the-art models and demonstrate its superior performance. The inference runs in an interactive interface that allows users to toggle enhancement, visualize spectrograms, and monitor processing latency.
\end{abstract}

\section{Introduction}
In extremely noisy environments, body-conducted microphones have been shown to outperform traditional air-conducted microphones in terms of speech intelligibility and quality \cite{hauret2023configurable}. However, since the human body acts as a natural low-pass filter, high-frequency components of speech are severely attenuated, and physiological noises such as heartbeat, breathing, and swallowing can contaminate the signal. This creates a clear need for post-processing speech enhancement to restore intelligibility and quality while preserving speaker identity \cite{joubaud2025french}.

While lightweight deep learning models have been proposed to enhance speech from non-conventional microphones \cite{hauret2023eben, yu2020time, sui2024tramba, ohlenbusch2025low}, they remain limited by the scarcity of real-world training data, as public datasets never exceed a hundred hours of recordings \cite{hauret2025vibravox}. Inspired by self-supervised learning strategies \cite{baevski2020wav2vec,hsu2021hubert,chen2022wavlm}, we propose leveraging neural audio codecs as foundation models for speech enhancement. These models, originally trained to approximate the identity function through a severe bottleneck, inherently learn clean speech patterns that can be repurposed for downstream enhancement tasks. Additionally, by operating on discrete token representations of speech rather than continuous waveforms, this paradigm reduces the dimensionality of the task while ensuring that the reconstructed output remains within the natural distribution of clean speech, borrowing the concept of "\textit{regeneration learning}" \cite{tan2024regeneration}.

Our demo builds on this paradigm by fine-tuning the encoder of a neural audio codec on paired throat and air-conducted recordings, enabling accurate reconstruction of full-band speech from body-conducted input. In contrast to prior work that stacks language models on top of acoustic tokens \cite{xue2024low, yang2024genhancer}, our approach retains the codec's simplicity. We posit that the effectiveness of our approach stems from the rich priors learned during large-scale pretraining, which help compensate for the spectral limitations inherent to body-conducted speech.

\section{Experimental setup \& Model selection}

We conduct experiments on Vibravox \cite{hauret2025vibravox}, a dataset of 38 hours of airborne and body-conducted speech sextuplets from 188 participants. We focus on the throat microphone for its superior noise robustness and use only the \verb+speech_clean+ subset (recorded in quiet conditions) as mixing with the \verb+speechless_noisy+ split did not yield significant gains in previous work \cite{hauret2025vibravox}. Three systems were trained until metric convergence:

\begin{itemize}
 \item \textbf{Mimi} \cite{defossez2024moshi}: the system showcased in this demo. We adopt a simple regression strategy where the encoder is duplicated: one is frozen to extract clean reference embeddings, while the other is initialized from the same checkpoint and fine-tuned to minimize the $\mathcal{L}_1$ distance between enhanced and reference embeddings.

 \item \textbf{EBEN} \cite{hauret2023configurable}: a lightweight model specifically designed for body-conducted speech enhancement. It shares architectural similarities with Mimi but omits sequence modeling layers to reduce complexity.

 \item \textbf{Nemo-FlowMatching} \cite{ku2025generative}: a flow matching foundation model available in NVIDIA NeMo \cite{kuchaiev2019nemo} \footnote{\url{https://huggingface.co/nvidia/sr_ssl_flowmatching_16k_430m}}. The original model was fine-tuned on bandwidth extension as one of its core subtasks, making it suitable for our setup. We retain the same fine-tuning and inference procedure as described in the original work. This model serves as a large-scale non-streamable state-of-the-art reference for comparison.
\end{itemize}

To evaluate the different approaches, we selected Noresqa-MOS \cite{manocha2021noresqa} and wideband STOI \cite{stoi}, both showing strong agreement with MUSHRA scores in previous body-conducted speech enhancement work \cite{hauret2023configurable}. We also included SI-SDR \cite{le2019sdr}, as an indicator of time-domain correlation and alignment between enhanced and reference signals, though less suited for assessing perceptual quality. Finally, we reported some Phoneme-Error-Rate (PER) between the reference phonemized transcription and the output of the airborne phonemizer \footnote{\url{https://huggingface.co/Cnam-LMSSC/phonemizer_headset_microphone}} fed with the enhanced audio. Results are indicated in Table \ref{tab:results}.

\begin{table}[th]
  \caption{Test set metrics of Vibravox \texttt{speech-clean}}
  \label{tab:results}
  \centering
  \begin{tabular}{l c c c c c}
    \toprule
    \textbf{Approach} & \textbf{Params} & \textbf{STOI} & \textbf{SI-SDR} & \textbf{N-MOS} & \textbf{PER} \\
    \midrule
    Throat                & -                & 0.677 & -7.99 & 3.10  & 50.8\% \\
    Mimi \cite{defossez2024moshi} & 96.2M            & 0.841 & 1.37  & 4.112 & 15.1\% \\
    EBEN \cite{hauret2023configurable}                        & 1.9M    & 0.834 & 3.21  & 3.862 & 18.6\% \\
    Nemo \cite{kuchaiev2019nemo} & 430M            & 0.822 & 7.30  & 4.398 & 7.6\%  \\
    \bottomrule
  \end{tabular}
\end{table}

\begin{figure*}[ht]
    \centering
    \begin{subfigure}[b]{0.42\textwidth}
        \centering
        \includegraphics[width=\textwidth]{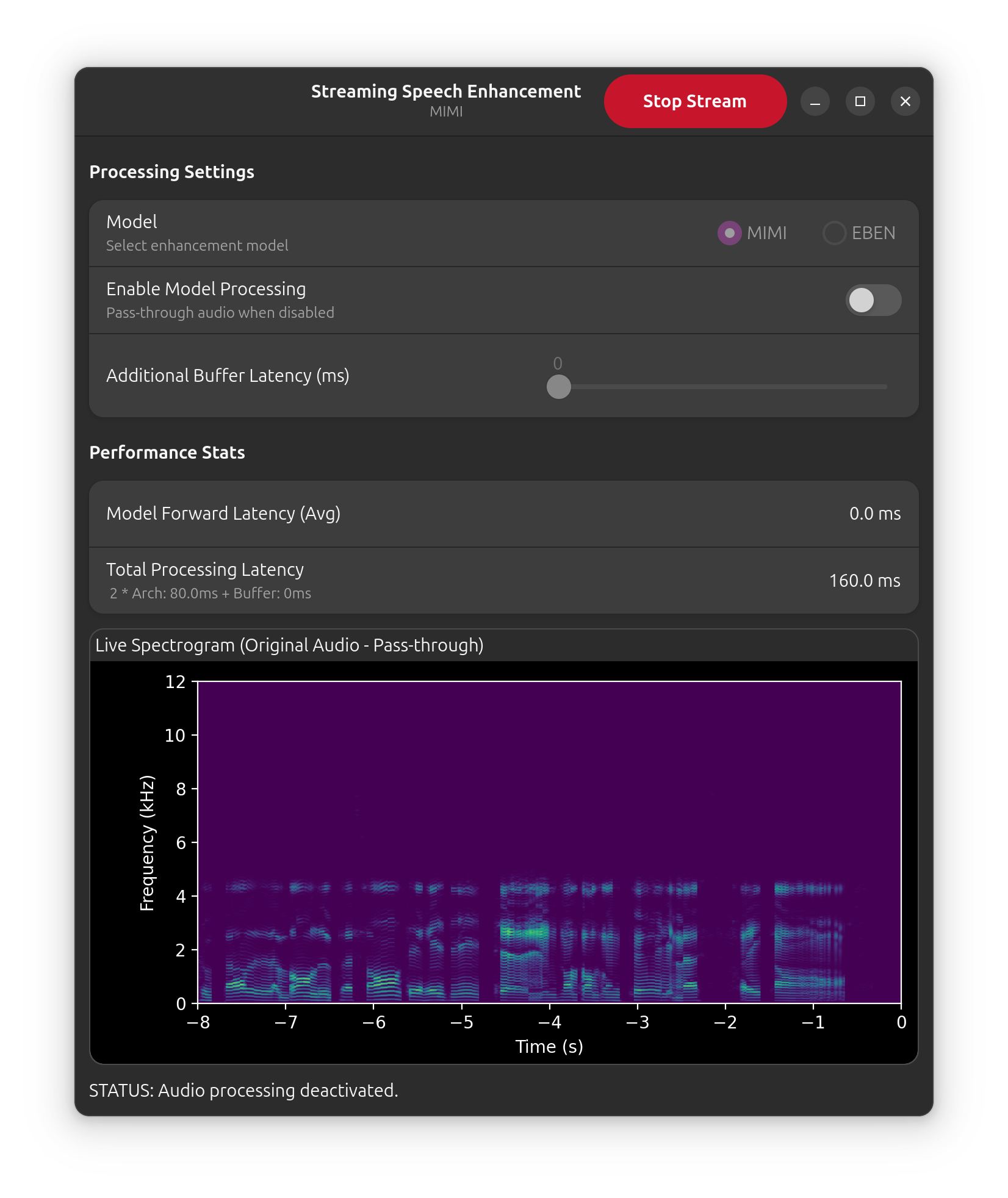}
        \caption{Without speech enhancement}
        \label{fig:processing_off}
    \end{subfigure}
    \hfill
    \begin{subfigure}[b]{0.42\textwidth}
        \centering
        \includegraphics[width=\textwidth]{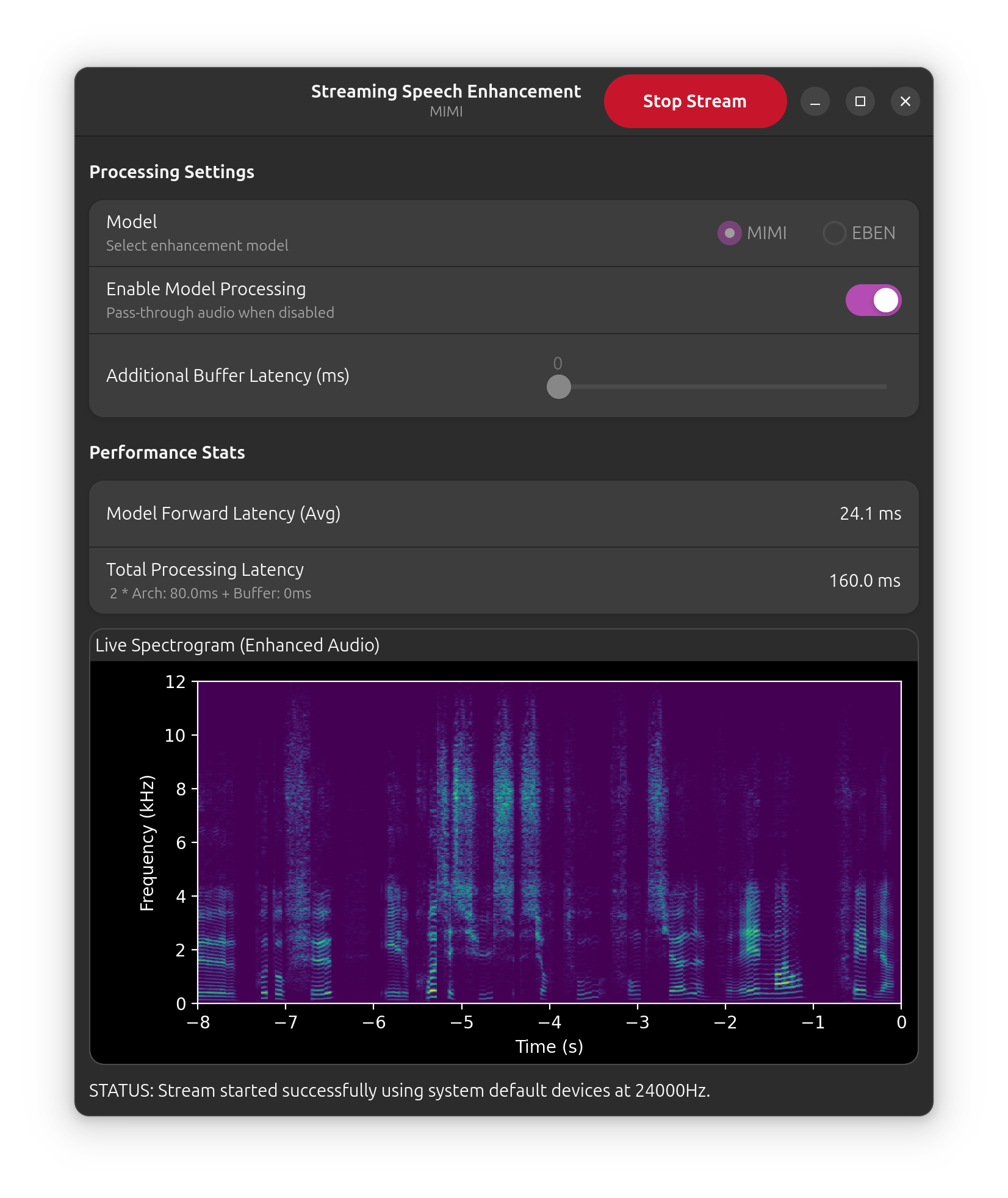}
        \caption{With speech enhancement}
        \label{fig:processing_on}
    \end{subfigure}
    \caption{User interface of the demonstration featuring speech captured with and without real-time enhancement.}
    \label{fig:enhancement_comparison}
\end{figure*}

From those results, we see that Mimi achieves the highest STOI score while maintaining a Noresqa-MOS comparable to the flow matching foundation model, but exhibits a lower SI-SDR. This decline in SI-SDR is due to the codec-based approach, which compresses audio into latent codes, sacrificing temporal coherence while maintaining perceptual quality. In rare cases, particularly with the most challenging speech segments, Mimi occasionally introduced minor phoneme hallucinations, hence a PER slightly below the FlowMatching model. Informal listening tests confirmed that both the codec-based approach and Nemo produce high-quality speech, with EBEN also performing strongly, though slightly below the other two. Overall, these findings are consistent with those reported in \cite{liu2024closer}, underscoring the reliability of the regressive fine-tuning and further validating the efficacy of neural audio codecs in serving as reliable foundation models for speech enhancement. These results, along with the fact that it supports real-time inference, made us select Mimi for the present demonstration.

\section{Demo Description}

\subsection{Hardware}
The demo employs an XVTM822D-D35 throat microphone, featuring dual piezoelectric transducers mounted on an adjustable neckband. This contact-based design captures vibrations from the skin near the vocal cords while effectively attenuating ambient noise. The microphone operates without external power and streams audio to a Linux laptop via a RØDE AI-Micro interface. Real-time enhancement is performed directly on-device, with the output accessible through a connected headset.

\subsection{Frontend}

As shown in Figure~\ref{fig:enhancement_comparison}, the demo includes a native Linux graphical interface built with GTK4\cite{gtk4} and Python. It displays a live spectrogram of either the raw or enhanced throat microphone signal, computed asynchronously in a separate worker thread to ensure smooth user interface performance. Users can toggle real-time enhancement, switch between models, and adjust buffer latency. Both model inference latency and total end-to-end system latency are updated live. The spectrogram is rendered using matplotlib\cite{matplotlib}, and processing can be enabled or disabled without interrupting the audio stream.

\subsection{Backend}

The backend is implemented in Python using PyTorch\cite{paszke2017automatic}, and handles real-time audio acquisition at 24kHz, enhancement, and playback. Audio input/output is managed by the Sounddevice library\cite{sounddevice}, using low-latency streams with fixed-size frames aligned to the model's frame rate. The Mimi codec processes audio by encoding each frame into discrete latent tokens and decoding them into full-band speech. Processing is performed on an NVIDIA GeForce GTX 1650 GPU with pre-warmed CUDA kernels to reduce startup latency. The end‑to‑end minimal latency is dominated by two 80 ms Mimi frames. In practice, additional input/output buffering on the RØDE AI‑Micro can add on the order of a few‑tens of milliseconds per side (we budget $\leq$32 ms each), for a total $\approx$ 160–224 ms. All inference runs in no-gradient mode, and responsiveness is maintained through careful resource management using Python’s contextlib and multithreading.

\section{Conclusion}

We present a fine-tuned neural audio codec serving as a foundation model for real-time speech enhancement from throat microphone recordings. Our results show that a simple regressive fine-tuning of the codec encoder yields competitive performance with a non-streamable large-scale flow matching model and outperforms the specifically body-conducted designed EBEN model. By using the clean speech patterns acquired on discrete latent representations, the system increases quality and intelligibility while enabling efficient real-time inference. The interactive demo confirms the practical viability of codec-based speech enhancement for body-conducted microphones and paves the way for audio codec usage for other data-scarce applications.


\clearpage
\bibliographystyle{IEEEtran}
\bibliography{../mybib}

\begin{thebibliography}{10}
\providecommand{\url}[1]{#1}
\csname url@samestyle\endcsname
\providecommand{\newblock}{\relax}
\providecommand{\bibinfo}[2]{#2}
\providecommand{\BIBentrySTDinterwordspacing}{\spaceskip=0pt\relax}
\providecommand{\BIBentryALTinterwordstretchfactor}{4}
\providecommand{\BIBentryALTinterwordspacing}{\spaceskip=\fontdimen2\font plus
\BIBentryALTinterwordstretchfactor\fontdimen3\font minus
  \fontdimen4\font\relax}
\providecommand{\BIBforeignlanguage}[2]{{%
\expandafter\ifx\csname l@#1\endcsname\relax
\typeout{** WARNING: IEEEtran.bst: No hyphenation pattern has been}%
\typeout{** loaded for the language `#1'. Using the pattern for}%
\typeout{** the default language instead.}%
\else
\language=\csname l@#1\endcsname
\fi
#2}}
\providecommand{\BIBdecl}{\relax}
\BIBdecl

\bibitem{hauret2023configurable}
J.~Hauret, T.~Joubaud, V.~Zimpfer, and {\'E}.~Bavu, ``Configurable {EBEN}:
  Extreme bandwidth extension network to enhance body-conducted speech
  capture,'' \emph{IEEE/ACM Transactions on Audio, Speech, and Language
  Processing}, vol.~31, pp. 3499--3512, 2023.

\bibitem{joubaud2025french}
T.~Joubaud, J.~Hauret, V.~Zimpfer, and {\'E}.~Bavu, ``French listening tests
  for the assessment of intelligibility, quality, and identity of
  body-conducted speech enhancement,'' in \emph{Interspeech}, Rotterdam,
  Netherlands, 2025.

\bibitem{hauret2023eben}
J.~Hauret, T.~Joubaud, V.~Zimpfer, and {\'E}.~Bavu, ``Eben: Extreme bandwidth
  extension network applied to speech signals captured with noise-resilient
  body-conduction microphones,'' in \emph{International Conference on
  Acoustics, Speech and Signal Processing (ICASSP)}.\hskip 1em plus 0.5em minus
  0.4em\relax IEEE, 2023, pp. 1--5.

\bibitem{yu2020time}
C.~Yu, K.-H. Hung, S.-S. Wang, Y.~Tsao, and J.-w. Hung, ``Time-domain
  multi-modal bone/air conducted speech enhancement,'' \emph{IEEE Signal
  Processing Letters}, vol.~27, pp. 1035--1039, 2020.

\bibitem{sui2024tramba}
Y.~Sui, M.~Zhao, J.~Xia, X.~Jiang, and S.~Xia, ``Tramba: A hybrid transformer
  and mamba architecture for practical audio and bone conduction speech super
  resolution and enhancement on mobile and wearable platforms,''
  \emph{Proceedings of the ACM on Interactive, Mobile, Wearable and Ubiquitous
  Technologies}, vol.~8, no.~4, pp. 1--29, 2024.

\bibitem{ohlenbusch2025low}
M.~Ohlenbusch, C.~Rollwage, and S.~Doclo, ``Low-complexity own voice
  reconstruction for hearables with an in-ear microphone,'' in
  \emph{International Conference on Acoustics, Speech and Signal Processing
  (ICASSP)}.\hskip 1em plus 0.5em minus 0.4em\relax IEEE, 2025, pp. 1--5.

\bibitem{hauret2025vibravox}
J.~Hauret, M.~Olivier, T.~Joubaud, C.~Langrenne, S.~Poir{\'e}e, V.~Zimpfer, and
  {\'E}.~Bavu, ``Vibravox: A dataset of french speech captured with
  body-conduction audio sensors,'' \emph{Speech Communication}, vol. 172, p.
  103238, 2025.

\bibitem{baevski2020wav2vec}
A.~Baevski, Y.~Zhou, A.~Mohamed, and M.~Auli, ``wav2vec 2.0: A framework for
  self-supervised learning of speech representations,'' \emph{Advances in
  neural information processing systems}, vol.~33, pp. 12\,449--12\,460, 2020.

\bibitem{hsu2021hubert}
W.-N. Hsu, B.~Bolte, Y.-H.~H. Tsai, K.~Lakhotia, R.~Salakhutdinov, and
  A.~Mohamed, ``Hubert: Self-supervised speech representation learning by
  masked prediction of hidden units,'' \emph{IEEE/ACM Transactions on Audio,
  Speech, and Language Processing}, vol.~29, pp. 3451--3460, 2021.

\bibitem{chen2022wavlm}
S.~Chen, C.~Wang, Z.~Chen, Y.~Wu, S.~Liu, Z.~Chen, J.~Li, N.~Kanda,
  T.~Yoshioka, X.~Xiao \emph{et~al.}, ``Wavlm: Large-scale self-supervised
  pre-training for full stack speech processing,'' \emph{IEEE Journal of
  Selected Topics in Signal Processing}, vol.~16, no.~6, pp. 1505--1518, 2022.

\bibitem{tan2024regeneration}
X.~Tan, T.~Qin, J.~Bian, T.-Y. Liu, and Y.~Bengio, ``Regeneration learning: A
  learning paradigm for data generation,'' in \emph{Proceedings of the AAAI
  Conference on Artificial Intelligence}, vol.~38, no.~20, 2024, pp.
  22\,614--22\,622.

\bibitem{xue2024low}
H.~Xue, X.~Peng, and Y.~Lu, ``Low-latency speech enhancement via speech token
  generation,'' in \emph{International Conference on Acoustics, Speech and
  Signal Processing (ICASSP)}.\hskip 1em plus 0.5em minus 0.4em\relax IEEE,
  2024, pp. 661--665.

\bibitem{yang2024genhancer}
H.~Yang, J.~Su, M.~Kim, and Z.~Jin, ``Genhancer: High-fidelity speech
  enhancement via generative modeling on discrete codec tokens,'' in
  \emph{Interspeech}, 2024, pp. 1170--1174.

\bibitem{defossez2024moshi}
A.~D{\'e}fossez, L.~Mazar{\'e}, M.~Orsini, A.~Royer, P.~P{\'e}rez,
  H.~J{\'e}gou, E.~Grave, and N.~Zeghidour, ``Moshi: a speech-text foundation
  model for real-time dialogue,'' \emph{arXiv preprint arXiv:2410.00037}, 2024.

\bibitem{ku2025generative}
P.-J. Ku, A.~H. Liu, R.~Korostik, S.-F. Huang, S.-W. Fu, and A.~Juki{\'c},
  ``Generative speech foundation model pretraining for high-quality speech
  extraction and restoration,'' in \emph{International Conference on Acoustics,
  Speech and Signal Processing (ICASSP)}.\hskip 1em plus 0.5em minus
  0.4em\relax IEEE, 2025, pp. 1--5.

\bibitem{kuchaiev2019nemo}
O.~Kuchaiev, J.~Li, H.~Nguyen, O.~Hrinchuk, R.~Leary, B.~Ginsburg, S.~Kriman,
  S.~Beliaev, V.~Lavrukhin, J.~Cook \emph{et~al.}, ``Nemo: a toolkit for
  building {AI} applications using neural modules,'' \emph{arXiv preprint
  arXiv:1909.09577}, 2019.

\bibitem{manocha2021noresqa}
P.~Manocha, B.~Xu, and A.~Kumar, ``Noresqa: A framework for speech quality
  assessment using non-matching references,'' \emph{Advances in neural
  information processing systems}, vol.~34, pp. 22\,363--22\,378, 2021.

\bibitem{stoi}
C.~H. Taal, R.~C. Hendriks, R.~Heusdens, and J.~Jensen, ``An algorithm for
  intelligibility prediction of time–frequency weighted noisy speech,''
  \emph{IEEE Transactions on Audio, Speech, and Language Processing}, vol.~19,
  no.~7, pp. 2125--2136, 2011.

\bibitem{le2019sdr}
J.~Le~Roux, S.~Wisdom, H.~Erdogan, and J.~R. Hershey, ``Sdr--half-baked or well
  done?'' in \emph{International Conference on Acoustics, Speech and Signal
  Processing (ICASSP)}.\hskip 1em plus 0.5em minus 0.4em\relax IEEE, 2019, pp.
  626--630.

\bibitem{liu2024closer}
A.~H.Liu, Q.~Wang, Y.~Gong, and J.~Glass, ``A closer look at neural codec
  resynthesis: Bridging the gap between codec and waveform generation,''
  \emph{arXiv preprint arXiv:2410.22448}, 2024.

\bibitem{gtk4}
``Gtk, version 4.0,'' \url{https://www.gtk.org/}, accessed: July 2025.

\bibitem{matplotlib}
J.~D. Hunter, ``Matplotlib: A 2d graphics environment,'' \emph{Computing in
  Science \& Engineering}, vol.~9, no.~3, pp. 90--95, 2007.

\bibitem{paszke2017automatic}
A.~Paszke, S.~Gross, S.~Chintala, G.~Chanan, E.~Yang, Z.~DeVito, Z.~Lin,
  A.~Desmaison, L.~Antiga, and A.~Lerer, ``Automatic differentiation in
  pytorch,'' in \emph{Conference on Neural Information Processing Systems},
  2017.

\bibitem{sounddevice}
\BIBentryALTinterwordspacing
M.~Geier \emph{et~al.}, ``Sounddevice,'' 2020. [Online]. Available:
  \url{https://python-sounddevice.readthedocs.io/en/0.3.15/index.html}
\BIBentrySTDinterwordspacing

\end{thebibliography}







\end{document}